# C++QEDv2

András Vukics

## Table of Contents

# Synopsis

C++QED is a framework for simulating open quantum dynamics in general. Historically, it has in the first place been developed for problems in moving-particle cavity QED, but since then has been applied in other fields as well. It is known to be able to simulate full Master equation up to several thousand, and quantum trajectories up to several hundred thousand dimensions.

The basic idea of the framework is to allow users to build arbitrarily complex interacting quantum systems from elementary free subsystems and interactions between them (below these are commonly referred to as "elements"), and simulate their time evolution with a number of available time-evolution drivers. Operating with elementary physical systems, the interface is of much higher level than in the popular Quantum Optics toolbox for Matlab [1], or the much less known, but venerable QSD (quantum state diffusion) C++ library [2]. In the latter two the interface is built on quantum operators, and general usage involves a considerable amount of boilerplate.

C++QEDv2 specifies a small grammar to describe composite quantum systems. This qualifies as a domain specific language (in this case embedded into C++), although, admittedly, this aspect should be further developed in the future.

Apart from providing a number of elements out of the box, there are several tools which facilitate the implementation of new elements. These are being added continuously, "on demand".

There exists a first release of the framework, which has been partially documented in a journal article [3]. Concerning physics this paper of course still applies, and the reader should consult it if something is not clear from the physics point of view in the following. The differences between C++QEDv1 and v2 are so numerous that it is easier to describe the similarities, which essentially lie in the basic idea of the interface as described above. Because of this, and despite all the differences, experience has shown that it is not difficult migrate elements and scripts written for v1 to v2. C++QEDv1 relied on a pure object-oriented design, while in v2 the emphasis is more on generic programming and template metaprogramming (TMP). In particular, the principal concept of the design of C++QEDv2 is that *all information available at compile time should be processed as such*, with the help of TMP. All in all, v1 can be regarded as a prototype of v2.

As of today, the following possibilities for time evolution are provided in the framework:

- **Full Master equation**

- **Single Monte Carlo wave-function** (MCWF) **trajectory**. We use a modification of the original method with higher order adaptive stepsize time evolution.

- **Ensemble of quantum** (at present, MCWF) **trajectories**. These are evolved serially at the moment, parallelization should be implemented here.

A number of other methods, like e.g. the quantum state diffusion, can be easily incorporated into the framework.

> **Note**
>
> In the case when the probability of quantum jumps vanishes, the MCWF method reduces to the simulation of the Schrödinger equation, so that the latter is naturally available in the framework.

# Performance issues

The framework is very sensitive to performance both in terms of computer resources and coding/design. In the latter aspect the goal, as always in software design, is to create maximally reusable code. Perhaps the most spectacular example is that the very same code, if it is written generally enough, can be used to calculate MCWF and full Master-equation evolution. In the former aspect, there are physical and computational methods to increase performance. Among the physical ones, the most notable is the maximal use of interaction picture, which may help to get rid of very separate timescales in the problem. Among the computational ones we can mention

- Maximal exploitation of special operator structures, i.e., sparse and tridiagonal matrices.

- The use of adaptive-stepsize methods for evolving ordinary differential equations (ODE).

- Judicious use of memory. The guideline is: If it is necessary to copy something, ask first whether it is *really* necessary.

In addition, we have to note that simulation of moving particles is inherently hard, since the Schrödinger equation is a partial differential equation, and we inevitably have to deal with both position and momentum representations, which are linked by Fourier transformation. In our problems, however, the particles are mostly moving in potentials created by light fields, mainly standing and running waves. In this case we can stay in momentum space during the whole time evolution, and no FFT is necessary [3].

A funny consequence is that in numerical physics the harmonic oscillator seems to be hard, while the cosine potential is easy.

# Installation

## Requirements

C++QEDv2 depends on a number of open source libraries:

- The **Boost C++ libraries** provide indispensable extensions to the C++ standard, and are *de facto* standard by their own right. The framework depends on a number of them, the most notable ones being Fusion, Lambda, MPL, Operators, and Preprocessor. On many systems (a selection of) the Boost libraries are available. They are packaged for Debian and Mac OS X. Alternatively, they can be downloaded and installed from the main Boost portal @ http://www.boost.org. Version 1.35 or higher is required.

- **GNU Scientific library (GSL)** provides a very wide variety of numerical solutions in a solid object-oriented design (in C!). They are not used directly, but are wrapped into C++ classes and functions, so that they are easily replaced. Packaged for Debian and Mac OS X, or can be downloaded from the GSL homepage @ http://www.gnu.org/software/gsl/. I haven't thoroughly determined the minimal version, but 1.5 is known to work.

These two are best installed on system level.

The following two libraries are stable, but under more or less steady development.

- **Blitz++** provides the fundamental data structure and hence performs a lot of numerics and is lying at the absolute heart of the framework. Blitz++ lives up to its name, as it provides near-Fortran performance in spite of the very high-level abstractions used in the library. This is achieved by TMP, which was discovered in prototype during the development of this very library. More on the Blitz++ homepage @ http://www.oonumerics.org/blitz/.

- The **Flexible Library for Efficient Numerical Solutions (FLENS)** is a very impressive effort to wrap BLAS-LAPACK functions in a high-level C++ interface. FLENS in turn depends on BLAS-LAPACK (ATLAS). More on the FLENS homepage @ http://flens.sourceforge.net/.

> **Important**
>
> At the time of this writing the released version of Blitz++ is version 0.9. Do not use this release! The CVS version is to be used instead. There is no knowing when a new release will appear, but that release can be used, of course.

At the corresponding websites instructions for installing the libraries can be found. The user can also opt for downloading the alternative C++QED package, which provides some help with the latter two libraries: execute the `getLibs.sh` script in the directory `C++Utils/thirdParty`. This will download the CVS versions of both libraries into the subdirectories `blitz` and `FLENS-lite`, respectively, and will also compile Blitz++. With FLENS, automatic configuration is not provided, but this may change in the future.

I will very much appreciate all feedback regarding also the installation of the framework. If there seem to be many problems with FLENS, I may also provide another alternative package which does not use FLENS at all. In turn, this will not contain those features that rely on eigenproblem calculations, such as assessing entanglement using the negativity of the partial transpose --- for sure, *I will not want to go back medling with LAPACK directly!*

# Compilation

The canonical way to compile the framework is the one using Boost.Build. This is best installed on system level. Typing `bjam` in the main directory will compile and link the whole framework, creating separate executables from the highest level programs residing in directory `scripts`. The default compilation mode is `debugging` mode, meaning that in this case a lot of runtime checks are compiled into the framework, which come from Blitz++, FLENS, and myself. Every time a new script is added it should be compiled and tested in this way because this can detect a *lot* of errors. When we are absolutely sure that everything is all right, for data collection we may compile with `bjam release`, in which all the checks are omitted and optimisations are used, making the programs *about an order of magnitude faster*.

> **Important**
>
> Maximum efficiency is achieved only if the framework is compiled with `bjam release`!

There is a `Makefile` which will automatically recognise the executables in directory `scripts`, compile the framework, and statically link it with necessary libraries. The support for this is, however, waning rapidly.

The content of the directory `C++Utils` is a small library of very diverse but quite general tools, that I have abstracted during the development of the framework, and use in several other projects. This may in time become a project on its own. The reader is encouraged to have a look in there, too: some modules may be useful.

C++QEDv2 has been successfully compiled on several Linux platforms and Mac OS X. In all cases the GNU C++ Compiler has been used. Portability to other compilers remains to be demonstrated.

# Writing and executing scripts

In the following we will cover how to use the framework on the highest level. The highest level is a C++ program of DSL-like grammar, which I like to call a *script*. Later it may be desirable to provide a Python frontend, or even a GUI.

A script creates an executable which defines and simulates a system of a particular layout. All information pertaining to the layout of the system is processed at compile time. Our compile-time algorithms can be regarded as C++ programs generating C++ programs in such a way that in the resulting executable all the compile-time information is encoded to yield a maximally efficient executable for the given layout.

A script is composed of a part in which the system is specified, and another, in which we do something with the system, most notably simulate its time evolution in one of the ways described in the Synopsis.

# An elementary example

The simplest case is when we want to simulate a free system alone. Assume that this free system is a mode of a cavity, which can be pumped and is lossy. We begin with defining the system, which is trivial in this case:

```
PumpedLossyMode mode(delta,kappa,eta,cutoff);
```

Once we have defined a system we can already use it for several things, e.g. to calculate the action of the system Hamiltonian on a given state vector, but since C++QED is a "framework for simulating open quantum dynamics", we will probably want to do something more.

Suppose we want to run a single MCWF trajectory. The system is started from a pure initial state, which is specified as

```
quantumdata::StateVector<1> psi(mode::coherent(alpha,cutoff));
```

that is, the mode is in a coherent state with amplitude `alpha`. `StateVector<1>` means that it is a state vector of a system featuring a single quantum number.

Next, we define our trajectory:

```
quantumtrajectory::MCWF_Trajectory<1> trajectory(psi,mode,...parameters...);
```

The first two parameters are clear, the only thing to note is that `psi` is taken as a reference here, so the initial `psi` will be actually evolved as we evolve the trajectory. A lot more parameters are needed, pertaining to the ODE stepper, the random number generation, and other things, but, as we will see below, the user will usually not have to worry about these.

All that remains is to run the trajectory, which is accomplished by

```
runDt(trajectory,time,dt);
```

This will evolve `trajectory` for time `time`, and display information about the state of the system after every time interval `dt`. What information is displayed is defined by the system. There is another version, which can be invoked like this:

```
run(trajectory,time,dc);
```

Here, `dc` is expected to be an integer, and it is the number of (adaptive) timesteps between two displays. Strange as this may seem, this version is actually more suited to the physics of the problem, since the timesteps will be small, when many things are happening, and then we want more output, too.

# Parameters

## Concept

In the above, the necessary parameters must be previously defined somewhere. Parameters can of course come from several sources, but an alternative I usually find most useful is to have sensible defaults for parameters, which can be overridden in the command line when we actually execute the program with a given set of parameters. This possibility is indeed supported by the framework. Consider the following program:

```cpp
#include "EvolutionHigh.h"
#include "Mode.h"

int main(int argc, char* argv[])
{
  ParameterTable p;

  ParsEvolution pe(p);         // Driver parameters
  mode::ParsPumpedLossy pm(p); // Mode parameters

  pe.evol=EM_MASTER;
  pm.cutoff=30;
  // ... other default values may follow

  update(p,argc,argv,"--"); // Parsing the command line

  // ****** ****** ****** ****** ****** ******

  mode::SmartPtr mode(maker(pm,QMP_UIP));

  mode::StateVector psi(init(pm));

  evolve(psi,*mode,pe);

}
```

This is a full-fledged script, so if you copy this into directory `scripts`, it will compile. Let us analyse it line by line. `ParameterTable` is the module from C++Utils which stores all the parameters of the problem, and enables them to be manipulated from the command line. It can store any type for which i/o operations are defined. Next we instantiate the actual parameters for the time-evolution driver(s) and the mode, respectively. All the modules in the framework provide corresponding `Pars` classes. In the following we specify the desired default values, e.g. we set that the default evolution mode should be Master equation [1]. Next, the command line is parsed by the `update` function. Parsing is not very sophisticated at the moment, but some errors *are* detected.

In the next line we instantiate our mode, but now instead of the concrete `PumpedLossyMode` class, we are using a maker (or dispatcher) function, which selects the best mode class corresponding to the parameters. There are 10 possibilities: `Mode`, `ModeSch`, `PumpedMode`, `PumpedModeSch`, `LossyMode`, `LossyModeUIP`, `LossyModeSch`, and `PumpedLossyMode`, `PumpedLossyModeUIP`, `PumpedLossyModeSch`. Roughly speaking, `mode::SmartPtr` is an entity that can store either of these classes. E.g. if `kappa=0`, and `eta=0`, then we will have a `Mode`; if `eta` is nonzero, a `PumpedMode`; and if both are nonzero, a `PumpedLossyMode`. The significance of this is that e.g. if the mode is not lossy, then the possibility of a quantum jump will not even be considered during time evolution, which speeds things up.

`Sch`, `UIP` and no suffix mean Schrödinger picture, unitary interaction picture, and "full" interaction picture, respectively. It is easy to see that if the system is not lossy, then the latter two coincide. Schrödinger picture is provided mostly for testing purposes, the performance is usually not optimal in this case.

> **Note**
>
> Master-equation evolution does not work with non-unitary interaction picture. The reason for this will be explained in the reference manual. Violation is detected at runtime.

What we are telling the maker function in the same line is that the picture should be unitary interaction picture. Alternatively, we could add this as a parameter as well, which can be achieved by putting the line

```cpp
QM_Picture& qmp=p.add("picture","Quantum mechanical picture",QMP_UIP);
```

anywhere between `ParameterTable` and `update`.

---

[1] Ultimate defaults are anyway given by the framework at the point where the `Pars` classes are defined, but these of course cannot always qualify as "sensible".

In the next line `mode::StateVector` is just another name for `StateVector<1>` (believe me, this *is* useful sometimes), and `init` is just another dispatcher, this time for the initial condition of the mode.

Finally, in the last line `evolve` is a dispatcher for different evolution modes and the two versions of `run`. So with this the evolution mode can be changed from the command line, e.g. depending on the dimension of the problem.

## Execution

If in the command line we specify the `--help` option, the program will display all the available parameters together with their types, a short description, and the default value. The names of the parameters are pretty much what you would expect. The type information becomes less and less readable for more and more complex types, so I am actually considering to remove this.

An example command line then looks like

```
MyFirstCppQedScript --eps 1e-12 --dc 100 --deltaC -10 --cutoff 20 \
                    --eta "(2,-1)" ...
```

There are some parameters that are "stronger" than others. E.g. if `--dc` is nonzero, then always the `run` version will be selected by the `evolve` function above, regardless of the value of `--Dt`. The latter will only be considered if `--dc` is zero, because in this case the `runDt` version will be selected. There is a similar relationship between `--minitFock` and `--minit`: the former will always override the latter.

> **Note**
>
> If `--evol ensemble` is selected, then always the `runDt` version will be used. I leave it as an exercise to figure out the reason for this.

# Example: a binary system

Imagine we would like to define a more complex system, in which a two-level atom (qbit) interacts with a single cavity mode with a Jaynes-Cummings type interaction. Both the qbit and the mode may be pumped, and they may also be lossy. First, we have to define the free elements of the system:

```
PumpedLossyQbit qbit(deltaA,gamma,etaA);
PumpedLossyMode mode(deltaC,kappa,etaC,cutoff);
```

or

```
qbit::ParsPumpedLossy pq(p);
mode::ParsPumpedLossy pm(p);
// ... update and whatever here
qbit::SmartPtr qbit(maker(pq,QMP_IP));
mode::SmartPtr mode(maker(pm,QMP_IP));
```

Here `qbit::maker` will dispatch exactly the same possibilities that we have seen for the mode above.

Next, we define the interaction between them:

```
JaynesCummings act(qbit,mode,g);
```

or

```
jaynescummings::Pars pjc(p);
// ... followed by
JaynesCummings act(qbit,mode,pjc);
```

> **Note**
>
> `JaynesCummings` is designed in such a way that it accepts not only concrete classes, but, bluntly speaking, anything that `qbit::SmartPtr` (`mode::SmartPtr`) can store. The same is true for all the interaction elements in the framework.

Now we have to bind together the two free subsystems with the interaction, which is simply accomplished by:

```
BinarySystem system(act);
```

In the case of a `BinarySystem` the complete layout of the system can be figured out from the single interaction element --- and this is trivial. `BinarySystem` is an extremely powerful module, the design of which reflects the basic idea behind the framework. It internally handles all the loops and slicing that are necessary to calculate e.g. the effect of the Hamiltonian of the qbit component if it is part of a binary system. It acts and feels like any other system, like, e.g., `Qbit` itself, the difference being that the latter has only one quantum number, while `BinarySystem` has two. A basic design principle of the framework is that it is fully recursive, that is, any composite system can act as an element of an even more complex system. [2]

Our next task is to define the initial condition:

```
StateVector<2> psi(qbit::state0()*mode::coherent(alpha,cutoff));
```

This is to say that the qbit is in its 0 state, and the mode is in a coherent state with amplitude `alpha`. Both states are of type `StateVector<1>`, meaning that they are state vectors featuring a single quantum number, and `*` means direct product of state vectors, so the result here is clearly a `StateVector<2>`. Direct product is *not commutative* in this case, and we have to comply with the order defined above for the free systems. Alternatively, we could have said

```
StateVector<2> psi(init(pq)*init(pm));
```

From this point on, usage is the same as we have seen above for the mode example. Since in this case the system is a `BinarySystem`, it will reach into its constituents for the informations to display, supplying either with the corresponding reduced density operator, which contains *all* information on the state of the subsystem.

If the system is not to be used for anything else, just for being `evolved`, we can shake off the burden of having to invent all these redundant names like `qbit`, `mode`, `act`, `system`, `trajectory`, and create everything *in place*. In this case a full-fledged script can be as terse as

---

[2] This can come in handy e.g. in the case of a complex atom, which has internal structure and motional degrees of freedom. These two are defined as separate elements, and we can use a `BinarySystem` to actually represent an atom with inner and outer degrees of freedom.

```cpp
#include "EvolutionHigh.h"
#include "Mode.h"
#include "Qbit.h"
#include "JaynesCummings.h"
#include "BinarySystem.h"

int main(int argc, char* argv[])
{
  ParameterTable p;

  ParsEvolution pe(p);

  qbit::ParsPumpedLossy pq(p);
  mode::ParsPumpedLossy pm(p);
  jaynescummings::Pars pjc(p);

  update(p,argc,argv,"--"); // Parsing the command line

  quantumdata::StateVector<2> psi(init(pq)*init(pm));
  evolve(psi,
         BinarySystem(JaynesCummings(maker(pq,QMP_IP),
                                     maker(pm,QMP_IP),
                                     pjc)),
         pe);

}
```

# Output of scripts

Following a header part, the time-dependent simulated data is displayed, organized into columns. The first two columns are time and timestep, respectively, and then, separated by tab characters, the data stemming from the different subsystems follows. A key to the data is provided in the header part of the output.

In the case of a single MCWF trajectory there is a last column, which, roughly speaking, shows how far the system was from a quantum jump to happen in the given timestep. If this number becomes negative, it means that a quantum jump has happened. In this case it would be easy to signal *which kind* of jump has happened, which is useful if e.g. the user wants to monitor fluorescence of a given atomic transition, or the output of a cavity.

The output can be piped into a file, or an output file can be specified with the `--o` option. In the latter case if the simulation comes to an end, the final state vector will be stored in a corresponding file with extension `.sv`. This allows the framework to resume a trajectory.

# More complex examples

If there are more than two free subsystems, the system can be much more complex. The number of possible interactions rises exponentially with the number of frees. This is the situation when the full potential of C++QED is displayed.

For the description of the elements appearing in the following examples cf. Ref. [3].

## Ring cavity

Assume we want to define a system in which a particle is moving along the axis of a ring cavity, and is interacting with two counterpropagating running-wave modes of the cavity. Both of the modes are lossy, and one of them is pumped; the particle is not pumped. This system consists of three subsystems, a particle, and the two modes. There are three interactions. (1-2) The particle can absorb from either of the modes and emit it in a stimulated way into the *same* mode. This yields dipole force for the particle and a corresponding light shift for the mode. It is implemented by the interaction element `ParticleAlongCavity`. (3) The particle can emit into the *other* mode. This yields a ternary interaction between all the frees, implemented by `ParticleTwoModes`.

We can lay out the system as the following simple network:

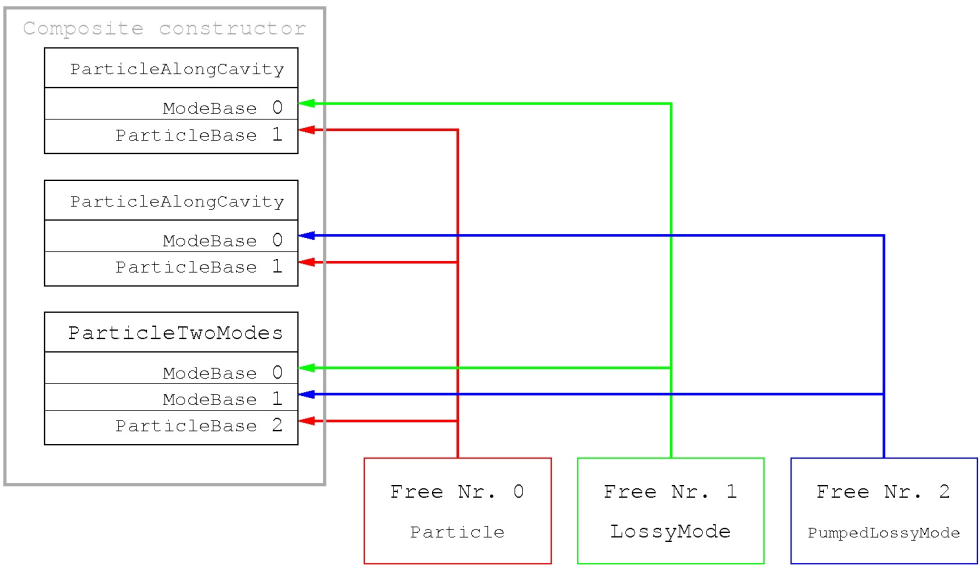

The `Composite` module of the framework is designed to represent such a network. Assume the following definitions are in effect:

```
// Instantiate Frees
Particle        part (...);
LossyMode       plus (...);
PumpedLossyMode minus(...);
// Instantiate Interactions
ParticleAlongCavity actP(plus ,part,...,MFT_PLUS );
ParticleAlongCavity actM(minus,part,...,MFT_MINUS);
ParticleTwoModes    act3(plus,minus,part,...);
```

Here `MFT_` means the type of the mode function and can be `PLUS`, `MINUS`, `COS`, and `SIN` [3].

Then the system can be created by invoking the maker function for `Composite` with a helper class, which, for some obscure reason, I decided to name `Act`:

```
makeComposite(
             Act<1,0>  (actP),
             Act<2,0>  (actM),
             Act<1,2,0>(act3)
             );
```

What we are expressing here e.g. with the specification `Act<1,2,0>(act3)` is that the 0th "leg" of the interaction element `ParticleTwoModes`, which is the mode `plus`, is the 1st in our row of frees in the network above.

> **Note**
>
> Following C/C++ convention, all ordinals begin with 0 in the framework.

The 1st leg, the mode `minus` is the 2nd in the row; and the 2nd leg, the particle is the 0th in the row of frees. The legs of interaction elements cannot be interchanged, and we also have to be consistent with our preconceived order of frees throughout. Clearly, the three `Act` objects above contain all the information needed by the framework to figure out the full layout of the system.

Any inconsistency in the layout will result in a compile-time or runtime error. I encourage the user to play around creating layout errors deliberately, and see what effect they have. Creating deliberate compilation errors as a response to misuse on a higher level, in such a way that the compiler is in addition *forced* to emit a sensible error message, is difficult. However, it is of course indispensable in template metaprogramming, if we want to leave any chance for ourselves to debug our metaprograms if something goes wrong. Here we are again relying on the Boost.MPL library.

The actual C++ type of a Composite object returned by such an invocation of `makeComposite` is quite complex, but for the sake of completeness we quote it here:

```
Composite<result_of::make_vector<Act<1,0>,Act<2,0>,Act<1,2,0> >::type>
```

Therefore, if we need a named object storing our Composite, we are better off with an additional `typedef`:

```
typedef result_of::make_vector<Act<1,0>,Act<2,0>,Act<1,2,0> >::type Acts;
Composite<Acts> system(Acts(actP,actM,act3));
```

A full-fledged script in the terse way may look as

```cpp
#include "EvolutionHigh.h"
#include "Composite.h"
#include "ParticleCavity.h"
#include "ParticleTwoModes.h"

int main(int argc, char* argv[])
{
  ParameterTable p;

  ParsEvolution pe(p);
  particle::Pars pp(p);
  mode::ParsLossy       pmP(p,"P");
  mode::ParsPumpedLossy pmM(p,"M");
  particlecavity::ParsAlong ppcP(p,"P");
  particlecavity::ParsAlong ppcM(p,"M");

  ppcP.modeCav=MFT_PLUS; ppcM.modeCav=MFT_MINUS;

  update(p,argc,argv,"--");

  particle::SmartPtr part (maker(pp ,QMP_IP));
  mode    ::SmartPtr plus (maker(pmP,QMP_IP));
  mode    ::SmartPtr minus(maker(pmM,QMP_IP));

  quantumdata::StateVector<3> psi(wavePacket(pp)*
                                  init(pmP)*
                                  init(pmM));
  evolve(psi,
         makeComposite(
                       Act<1,0>  (ParticleAlongCavity(plus ,part,ppcP)),
                       Act<2,0>  (ParticleAlongCavity(minus,part,ppcM)),
                       Act<1,2,0>(ParticleTwoModes(plus,minus,part,ppcP,ppcM))
                       ),
         pe);

}
```

A notable additional feature as compared to previous examples is that since now we have two modes in the system, we somehow have to differentiate between their parameters in the command line. This is achieved by the `"P"` and `"M"` modifiers added to the constructors of `Pars` objects, so that e.g. instead of `--cutoff` we now have the separate options `--cutoffP` and `--cutoffM`.

Although now all the frees have the general types contained by the `SmartPtr` classes, their possible types are still restricted by the `Pars` classes, such that e.g. `plus` can never become pumped.

Most of the interactions in C++QED will be binary, here we have seen an example for a ternary interaction. I know of only a single example for a four-leg interaction.

# Self-organisation example

Finally we are reviewing one more example, which displays a last feature, which, in turn, reflects a basic principle of quantum physics: if two systems are identical, they are indistinguishable. In the language of C++QED this means that a single object is enough to represent them.

Consider two identical pumped particles moving in a direction orthogonal to the axis of a cavity sustaining a single lossy mode. The layout of the system is:

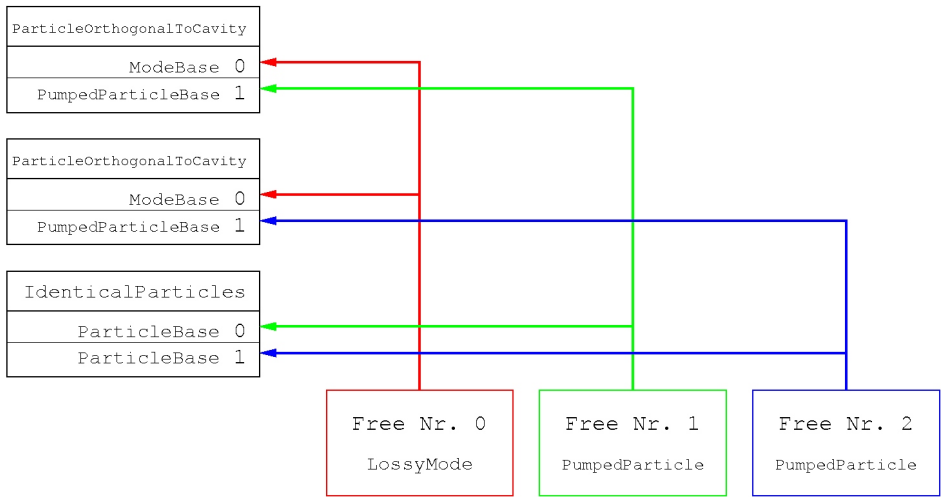

Without much ado we are quoting the kernel of a corresponding script:

```
LossyMode       mode(pm); // Free0
PumpedParticle part(pp); // Free1,2 - only one instant

ParticleOrthogonalToCavity act(mode,part,ppc); // only one instant

quantumdata::StateVector<3> psi(init(pm)*coherent(pp)*coherent(pp));

evolve(psi,
       makeComposite(
                Act<0,1>(act),Act<0,2>(act),
                Act<1,2>(IdenticalParticles<2>(part,...))
                ),
       pe);
```

(A pumped particle can also be in a coherent state: a coherent state of the pump potential approximated as harmonic.)

# Assessing entanglement

In a composite system we may want to assess the entanglement between two parts of the system. This can be done using the negativity of the density operator's partial transpose. Of course, since the dependence of this quantity on the density operator is not linear, this makes sense only in the case of Master-equation evolution or an ensemble of quantum trajectories.

The subsystem to be considered as one party of the two has to be specified in an additional compile-time vector argument to the `evolve` function.

To show the syntax we assume e.g. that in the previous example we are looking for the entanglement between the two particles together as one party, and the mode as the other party. Then the invocation of `evolve` is modified as

```
evolve(psi,system,pe,
       tmptools::Vector<1,2>());
```

We simply have to list in my compile-time vector the frees that consist one party. Of course in this case this is equivalent to `tmptools::Vector<0>`. Later I may invent a better name for the vector when used for this special purpose.

The negativity will appear as a last column in the output, separated by a tab character from the rest.

# Release

The current release of the framework is C++QEDv2 Milestone 8, and it is a bugfix release. The development is now in beta stage with no known major bugs. The foreseeable steps in the development are as follows:

- **Milestone 9** will concentrate on improving the documentation. A reference or extenders' manual will be provided.

- **Milestone 10** will be a release with some new features implemented. For some of these, the idea arose when writing this tutorial.

- **Milestone 11** will be a release with improved coding. While the runtime performance should be close to optimal by now, there is probably a lot to improve in the template and preprocessor metaprogramming parts. This can significantly cut on resources needed to compile the framework.

- **Milestone 12** will see the possibility to use non-orthogonal bases for free elements implemented. The framework is already prepared for this. A prominent example is of course modes in coherent-state bases.

- **Milestone ...** will achieve complete recursiveness in the definition of composite systems.

- **Milestone ...** will see the creation of a more general quantum-operator class of which `Tridiagonal` will be only one implementation, while others can be operators with sparse and full matrices. They should be arbitrarily combinable with expression-template like closures taking care of the necessary internal loops.

# Note on support

I offer full support for the framework both in terms of writing and testing new elements and scripts on demand from the quantum optics community, and of advising with the use of the existing software in the framework or with the development of new software (elements, scripts, time-evolution drivers). In the first case I may require to become co-author in the publications stemming from the work. In the second case I will probably only ask to please cite our first C++QED paper. I ask the same from anybody using C++QED without my support.

# Acknowledgement

First of all I would like to acknowledge the developers of Boost for making C++ an even more powerful and fun-to-work-with language than it originally was. Without the Boost libraries, the framework could not have been achieved by a single person.

I would like to sincerely thank the developers of GSL, LAPACK, Blitz++, and FLENS for their absolutely fantastic effort, without which scientific computing in C++ in general, and the present framework in particular would not look as nice as it looks today.